\documentclass{aa}
\usepackage{graphicx}
\usepackage{amsmath}
\usepackage{natbib}
\bibpunct{(}{)}{;}{a}{}{,}
\begin{document}

\title{On the Eddington limit and WR Stars}
\titlerunning{The Eddington limit}

\author{Andr\'e Maeder $^1$ \and Cyril Georgy $^2$  \and Georges Meynet $^1$ \and Sylvia Ekstr\"{o}m $^1$}
\authorrunning{Maeder et al.}

\institute{$^1$Geneva Observatory, Geneva University, CH--1290 Sauverny, Switzerland, 
              email: andre.maeder@unige.ch, georges.meynet@unige.ch, sylvia.ekstrom@unige.ch\\
              $^2$Centre de Recherche Astrophysique, Ecole Normale Sup\'erieure de Lyon, 46, all\'ee d'Italie, F-69384 Lyon cedex 07, France, email:cyril.georgy@ens-lyon.fr
}

\date{Received  / Accepted }

\offprints{Andr\'e Maeder} 
   
\abstract{}{We examine some properties of stars evolving close to the classical Eddington limit for electron-scattering opacity, when these stars maintain a chemically homogeneous structure as a result of mixing and/or mass loss.}{We consider analytical relations and models computed with the Geneva code.}{Homologous, chemically  homogeneous stars evolving with a constant Eddington factor obey a relation of the form $\mu^2 M = \mathrm{const}$. This  applies, for example, to Wolf-Rayet (WR) stars in stages without hydrogen. The value of the constant may depend on the metallicity, initial mass, evolutionary stage, and physical processes included in the considered homologous evolutionary sequence. An average value of the constant between 20 and 40 in solar units is consistent with the masses of Galactic WR stars.}{}


\maketitle
\section{Introduction}

Stars must have a luminosity inferior to the  Eddington limit, otherwise ``\textit{... the radiation observed to be emitted ... would blow up the star}'' as written by \citet{Edd26}. This limit, which occurs when the outward acceleration of radiation pressure becomes equal to gravity, usually has a simple analytical expression (see Eq. \ref{ledd}). Here, we present another simple expression of the Eddington limit applicable to a sequence of homologous stellar models that evolve without losing their homogeneous chemical structure. 

During their evolution, stars generally become chemically inhomogeneous, hence our proposed expression would not be applicable. However, in some relatively rare cases, stars can indeed maintain a chemically homogeneous structure during the H- and He-burning phases. This may result in massive stars undergoing a heavy mass loss that removes the outer stellar layers \citep{Maeder80}, or in strong internal mixing \citep{Maeder87b}. 

The latest grids of stellar models developed by the Geneva group \citep{Ekstrom2012a} show homogeneous evolutionary tracks for stellar masses above 50 $M_{\sun}$ with even moderate initial rotation at 40\% of the critical velocity. This confirms that depending on rotational mixing a bifurcation may occur in the general scheme of stellar evolution: tracks evolve toward the blue instead of the usual way toward the red. For more rapid rotation, homogeneous evolution appears above lower mass limits.The homogeneous stars rapidly enter the Wolf--Rayet (WR) stages as shown by  \citet{mm11} and \citet{Georgy2012b}. 

The present work on the Eddington limit of homogeneous stars is in line and consistent with the results of \citet{graefener11}  showing that "\textit{the proximity to the Eddington limit is the physical reason for the onset of WR type mass loss}". The  expression we propose here may be particularly useful for deriving the mass loss rates of WR stars.

An additional motivation for this study is that the homogeneous evolution of massive hot stars has been proposed as the scenario leading to long--soft gamma ray bursts \citep{YL05}. In this case, homogeneity would be due to the rapid axial rotation of the star. Homogeneous stars can also be formed during stellar merging. This scenario was reexamined by \citet{glebbeek09} as a possible means of forming intermediate-mass black holes in globular clusters.
 
In Sect. \ref{UppLim}, we proceed to some basic analytical developments. In Sect.~\ref{SecPoly}, we consider a polytropic structure and in Sect.~\ref{SecHomo} the case of homologous evolution. WR observations are examined in Sect.~\ref{SecObs}, while Sect.~\ref{SecConclu} presents our conclusions.

\section{The upper mass limit for  homogeneous stars\label{UppLim}}

\subsection{The mass-luminosity relation}

We begin with and develop briefly some basic relations. The equation of radiative equilibrium
\begin{equation}
\frac{L_r}{4\pi r^2} =- \frac{ac}{3\kappa\rho}\frac{\mathrm{d}T^4}{\mathrm{d}r},
\end{equation}
\noindent
implies the scaling for the luminosity $L$
\begin{equation}
L \propto \frac{R}{\kappa \rho} T_\mathrm{c}^4,
\end{equation}
where $\kappa$  and $\rho$ are the mean values of the opacity and density in a star of radius $R$ and central temperature $T_\mathrm{c}$. The equation of hydrostatic equilibrium $\frac{\mathrm{d}P}{\mathrm{d}r}=-\rho g$ gives a scaling for the central pressure
\begin{equation}
P_\mathrm{c} \propto \frac{M^2}{R^4}.
\label{pc}
\end{equation}
The equation of state for massive hot stars is that of a mixture of perfect gas and radiation $P={P_\mathrm{gas}}/{\beta}={k} \rho T/ {(\beta \mu m_\mathrm{u}})$, where $\beta$ is the ratio of the gas pressure to the total pressure. This implies that
\begin{equation}
T_\mathrm{c} \propto \frac{\mu \beta_\mathrm{c}}{\rho_\mathrm{c}} P_\mathrm{c} 
\propto \frac{\mu \beta_\mathrm{c}}{(\rho_\mathrm{c}/\rho)}\frac{M}{R}.
\end{equation}
As the star is chemically homogeneous, the mean molecular weight $\mu$ is assumed to be constant throughout, (for cool stars $\mu$ may vary owing to changes in the degree of ionization).  For this expression of $T_{\mathrm{c}}$, the luminosity $L$ behaves like
\begin{equation}
L  \; = \;  A  \mu^4 \beta_\mathrm{c}^4 \frac{\rho^4}{\rho_\mathrm{c}^4}\frac{M^3}{\kappa},
\label{lum}
\end{equation}
where $A$ is a proportionality constant. The above scalings  are rather rough, thus $A$ is almost constant only as long as the model structures do not differ too considerably. We consider below the polytropic and homologous structures. The term $\beta$ and the ratio of the two densities are often omitted from the mass-luminosity relation. 
\begin{figure}[t]
\begin{center}
\includegraphics[width=9.0cm, height=7.5cm]{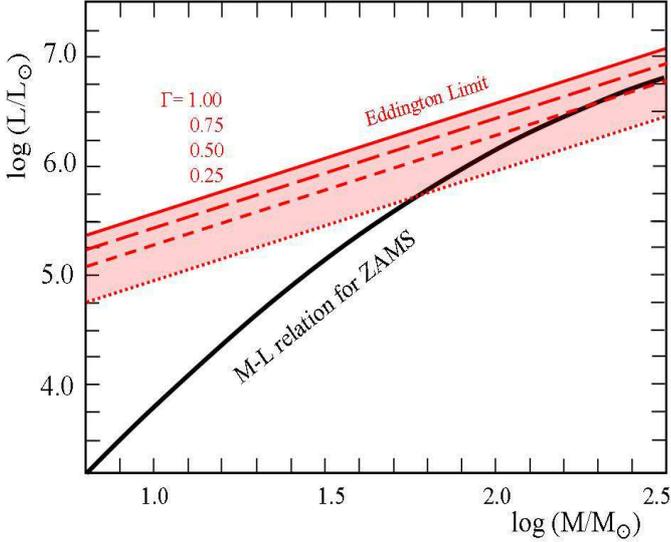}
\caption{Mass--luminosity relation for ZAMS stars with $X=0.70$ and $Z=0.014$ \citep[from][in black]{graefener11}, compared to the limits given by Eq.~\ref{lg} for the same composition and various $\Gamma$--values as indicated.}
\label{eddzams}
\end{center}
\end{figure}

\subsection{The Eddington limit}

According to the definition, the Eddington luminosity is given by
\begin{equation}
\frac{1}{\rho_\mathrm{s}}\frac{\mathrm{d}P_\mathrm{rad}}{\mathrm{d}r}= \frac{\kappa_\mathrm{s}}{c}\frac{L_\mathrm{EDD}}{4\pi R^2}=\frac{GM}{R^2} \; .
\end{equation}
The first equality expresses the gradient of the radiation pressure in term of the radiative flux, while the second says that this gradient is equal to the modulus of gravity. The subscript "s" stands for quantities at the surface. This gives the classical Eddington limit
\begin{equation}
{L_\mathrm{EDD}}=\frac{4\pi cGM}{\kappa_\mathrm{s}} \; .
\label{ledd}
\end{equation}
Here, it is the opacity  $\kappa_\mathrm{s}$ in layers close to the surface which intervenes. However, in hot massive stars, the main opacity source is electron scattering. To first order, the opacity is the same throughout a homogeneous star, thus here we no longer distinguish the two $\kappa$ values.

The maximum luminosity effectively observed $L_{\Gamma}$ may lie below the formal limit $L_{\mathrm{EDD}}$ and be given by
\begin{equation}
L_{\mathrm{\Gamma}} \;= \; \Gamma\; L_{\mathrm{EDD}} \; ,
\label{lg}
\end{equation}
which is the effective limit for some $\Gamma < 1$. 
 
We consider the possibility that the real $\Gamma$ is lower than 1.0, since there are several physical reasons for that. Firstly, the actual opacities in the outer stellar layers are higher than just the electron-scattering opacities assumed for the classical Eddington limit. In particular, strong  resonance spectral lines are driving the heavy radiative mass loss of massive stars. Any pulsational instabilities, that is either radial or non-radial oscillations, are favoured by the decrease in the adiabatic exponents $\Gamma_1$, $\Gamma_2$ and $\Gamma_3$. These exponents are all reduced by an increase in the ratio of the radiation pressure to the gas pressure, as it occurs close to the Eddington limit. The excitation mechanisms of either p modes or g modes, as observable in asteroseismology, are  also  enhanced by a high radiation pressure. This is also the case for the strange modes developing in very massive stars with a high radiation pressure \citep{papaloizou97}. Thus, there are several effects and instabilities that may place the real maximum luminosity of stars of a given mass and composition below the academic value of the Eddington limit. These theoretical considerations are in close agreement with observations. For example,  all the most luminous WR stars in the Arches cluster have $\Gamma$--values between 0.2 and 0.45 \citep{graefener11}, i.e. at some significant distance below the formal limit.

We emphasise that the effective $\Gamma$ may depend on the parameters of the considered stellar population, such as its metallicity $Z$, etc. For example, for a higher metallicity $Z$, the actual opacity in the innermost layers of the wind is larger, implying that the momentum of radiation onto the outer stellar layers is larger, and also that pulsational instabilities due to the $\kappa$--mechanism are enhanced. Thus, the effective $\Gamma$ is likely to be lower at higher $Z$. Other effects influencing the mass loss, such as rotation or tidal effects in binary system, may also influence the effective $\Gamma$.

\begin{figure}[t]
\begin{center}
\includegraphics[width=9.0cm, height=7.5cm]{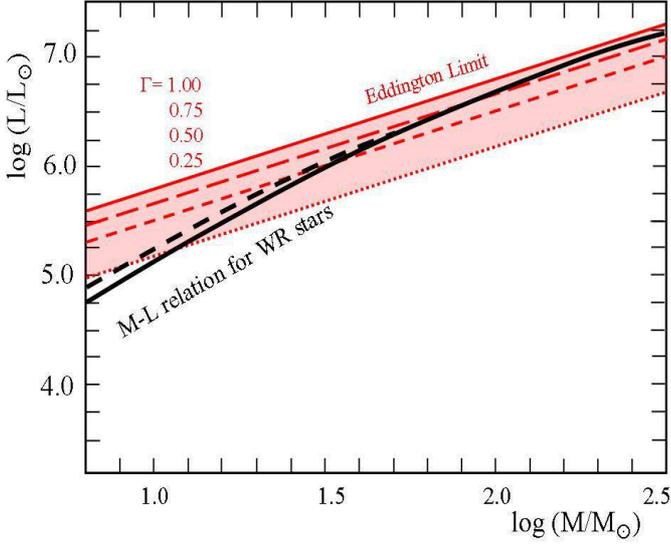}
\caption{Mass--luminosity relation for WR stars without hydrogen (in black) compared to the limits given by Eq.~\ref{lg} for the same composition and various $\Gamma$--values as indicated. The continuous black line refers to models by \citet{graefener11} with $X=0$, while the broken black line represents the average mass-luminosity of WR stars without hydrogen from \citet{SchaerMaed92}.}
\label{eddwr}
\end{center}
\end{figure}

\subsection{The effective limit}

We assume that the highest stellar luminosity $L$ given by the mass-luminosity ($M-L$) relation \ref{lum} is equal to $L_{\mathrm{\Gamma}}$ defined by Eq.~\ref{lg},
\begin{equation}
A  \mu^4 \beta_\mathrm{c}^4 \frac{\rho^4}{\rho_\mathrm{c}^4}\frac{M^3}{\kappa}=
\Gamma\; \frac{4\pi cGM}{\kappa_\mathrm{s}} \; .
\label{ll}
\end{equation}

Figures \ref{eddzams} and \ref{eddwr} show the lines corresponding to various $\Gamma$ and the mass-luminosity relation from numerical models, respectively, for stars on the zero-age main sequence (ZAMS) and for WR stars without hydrogen. In both cases, we see that for $\Gamma=1$ the $M-L$ relation becomes parallel to the limit for high masses, implying that the limit is never reached, or in other words is only reached for extremely high masses or a value of $\beta$ tending toward zero. Therefore, Eq.~\ref{ll} for $\Gamma=1$  would not correspond to realistic cases. The factor $\Gamma$ measures the proximity to the Eddington limit. For  $\Gamma$--values lower than 1.0, the $M-L$ relation may intersect a $L_{\mathrm{\Gamma}}$--line  for some realistic stellar masses  $M_{\Gamma}$. For example, the line $\Gamma=0.25$ intersects the $M-L$ relation for a star of about 57 $M_{\sun}$ on the ZAMS (Fig. \ref{eddzams}), while for stars without hydrogen the same $\Gamma$-value  is reached by a star of about  12 $M_{\sun}$ (Fig. \ref{eddwr}).

Relation (\ref{ll}) defines the maximum mass $M_{\Gamma}$ allowed for a given value of $\Gamma$
\begin{eqnarray}
 \quad \mu^2 \,M_{\Gamma}   =  \Gamma^{1/2}\left( \frac{4 \pi \,c  \,G }{ A} \right)^{1/2}\,\, \left(\frac{\rho_\mathrm{c}}{\rho}\right)^2
\, \, \left( \frac{1}{\beta_{\mathrm{c}}} \right)^2 \; , 
\label{mucarre1}
\\
\mathrm{with\; A \; \; in \; \; cm^4\; s^{-3}\; g^{-3} \; ,} \nonumber  \\
\mathrm{with} \; \left( \frac{4 \pi \,c  \,G }{ A} \right)^{1/2}  \; \mathrm{in \; g} \; , \nonumber
\end{eqnarray}
for chemically homogeneous stars with electron-scattering opacity. The first parenthesis on the right-hand side is a constant. We see that in general the product $\mu^2 \,{M}$ is limited by a product of terms,  where the last two are not  constant. This product is examined in the next few sections.

\section{The polytropic case $n = 3$\label{SecPoly}}

We consider equation (\ref{mucarre1}) for some particular models. We first examine the case of a polytropic structure with a pressure of the form $P=K\, \varrho^{1+1/n}$ and index $n=3$. If internal variations in $\beta$ could be neglected, this case would be a good approximation for stars with an equation of state of perfect gas and radiation, since  the pressure $P$ could be written
\begin{equation}
P \, = \,  \left[\frac{3}{a} \left(\frac{k}{\mu \, m_{\mathrm{u}}}\right)^4
\, \frac{1-\beta}{\beta^4} \right]^{\frac{1}{3}} \, \varrho^{\frac{4}{3}} \; ,
\label{pgpr}
\end{equation}
 For such a polytrope, the equilibrium equations shows that the ratio of a star of mass $M$ to the maximum mass $M_{\mathrm{max}}$  for this type of objects is \citep{Edd26}
 \begin{equation}
 \left(\frac{GM}{M_{\mathrm{max}}}\right)^2 = \frac    {(4 \,K)^3}{4 \,\pi \, G} \; .
 \end{equation}
 With the expression of $K$ given by the bracket term with power $1/3$ in Eq.~(\ref{pgpr}), this becomes 
 \begin{equation}
 1-\beta= \left(\frac{ 4 \pi\, G^3 \,a \, {m^4_{\mathrm{u}}}}{3 \cdot 4^3 \, k^4 {M^2_{\mathrm{max}}}} \right) M^2 \, \mu^4 \beta^4 \; ,
 \end{equation}
The parenthesis contains only constant terms and is equal to $7.562 \cdot 10^{-70}$ g$^{-2}$, where we have taken $M_{\mathrm{max}}=2.018$ \citep{Chandra39}.

\begin{figure}[t]
\begin{center}
\includegraphics[width=9.0cm, height=7.0cm]{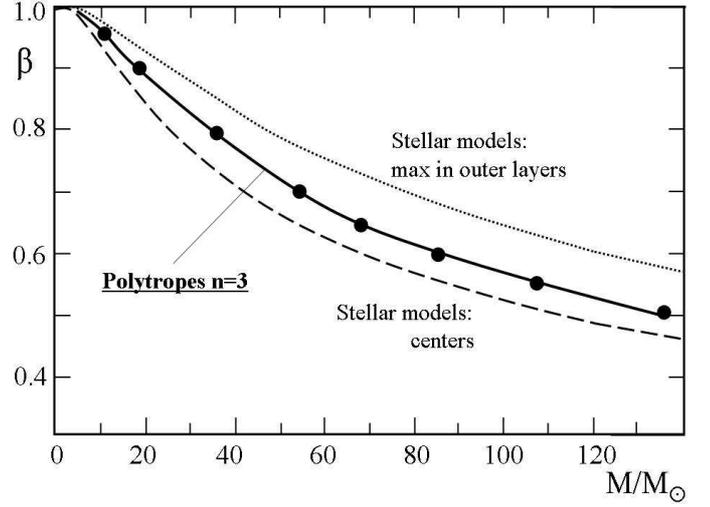}
\caption{Variation of $\beta$ as a function of stellar mass for models on the ZAMS with $X=0.705$ and $Z=0.02$. The maximum in outer layers and the minimum value in the centre for a given mass are shown \citep{MaederBook09}. The polytropic case given by Eq.~\ref{functionBmu} lies between the two curves.}
\label{betafmass}
\end{center}
\end{figure}

\begin{figure}[ht]
\begin{center}
\includegraphics[height=7.0cm,width=9.0cm]{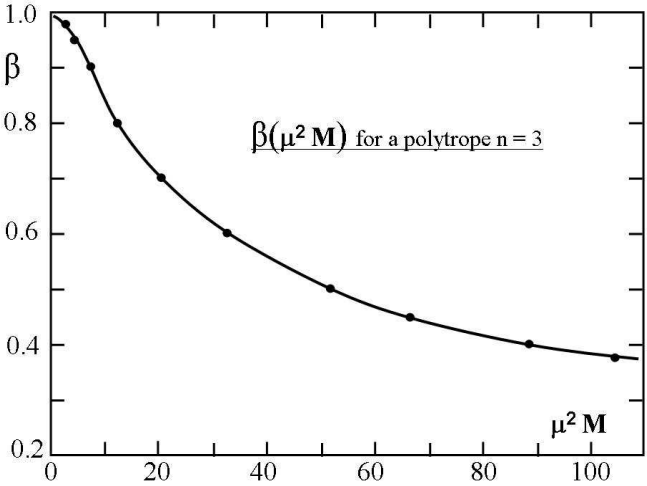}
\caption{The relation expressing $\beta$ as a function of  $\mu^2  M$ for a polytrope with index $n=3$, as given by Eq.~\ref{functionBmu}, where $M$ is expressed in solar units.}
\label{betapolytr}
\end{center}
\end{figure}

Expressing the mass  in solar units, we find that
\begin{equation}
\frac{(1-\beta)^{1/2}}{\beta^2} \, = \,  0.0547 \,\, \mu^2 \, (M/M_{\sun})  \; .
\label{functionBmu}
\end{equation}
This variation in the parameter $\beta$ as a function of  $M$ is illustrated in Fig. \ref{betafmass}, where it is also compared  to the central and external values of realistic stellar models on the ZAMS. The above relation indicates that, in a polytrope, the ratio $\beta$ of the gas to the total pressure is  a function only of $\mu^2 M$. This function is illustrated in Fig.~\ref{betapolytr}. 

In a polytrope with a given index $n$, the ratio of the central to the average density is always the same. For  $n=3$, the ratio $({\rho_\mathrm{c}}/{\rho})$ is equal to 54.2. Thus, besides the term $\Gamma^{1/2}$, the right-hand side of Eq.~\ref{mucarre1} contains only constant terms and a function $\beta=\beta(\mu^2 M)$. This function is monotonically decreasing (Fig. \ref{betapolytr}).  For high masses (say above 80 $M_{\sun}$), the parameter $\beta$ tends to vary as $(\mu^2  M)^{-1/2}$. Thus, the right-hand side of Eq.~\ref{mucarre1}, which contains the term $(1/\beta)^2$, behaves for very high masses like $\mu^2 \, M$. Thus,  Eq.~\ref{mucarre1} has no solution in this limit. This is the same limiting situation as for the usual Eddington limit with $\Gamma=1$ illustrated in Figs. \ref{eddzams} and \ref{eddwr}.

However, for not too large values of  $\mu^2  M$, the function $\beta^{-2}$, which appears on the right-hand side of Eq.~\ref{mucarre1}, grows faster than for very high masses. Thus, for some ranges of $\Gamma$--values smaller than 1.0, the curves representing the left and  right--hand sides of Eq.~\ref{mucarre1} may cross each other. This defines the maximum possible mass $M_{\Gamma}$ for the considered $\Gamma$--values. 

The solution also depends on the value of the parameter $A$ in the mass-luminosity Eq.~\ref{lum}. This relation is not defined by the polytropic structure, which just results from the mechanical equilibrium expressed by Poisson's equation. Nevertheless, the polytropic case is extremely interesting, since it shows that the product $\mu^2 M$ may be bounded.

\section{Homology relations\label{SecHomo}}

The polytropic case is rather restrictive and we now consider the more general case of an homologous sequence of chemically homogeneous models. These are models in which some parameters, such as the density and linear dimensions,  are multiplied at each point by some  constant  factors to produce another equilibrium configuration \citep{Chandra39}. This typically correspond to a sequence of models with similar structures that differ by a slowly varying parameter. Here, the changing parameter is the mean molecular weight $\mu$ in an evolutionary  sequence of chemically homogeneous stars. We examine the implications of the  limit  given by Eq. (\ref{mucarre1}) for such a sequence. 
 
 First, in a homologous sequence of stellar models with the same density profile, the ratio $({\rho_\mathrm{c}}/{\rho})$ remains constant. We now have to examine the behaviour of $\beta$. We make a homology transformation $r = \mathrm{constant}$ and $M'_{\mathrm{r}}= x \, M_{\mathrm{r}}$. The equation of hydrostatic equilibrium $dP/dr = - (GM_{\mathrm{r}}/r^2) \, \varrho$ implies that
 \begin{equation}
 P' \, = x^2 \, P \; .
 \end{equation}
Thus, in Eq.~\ref{pgpr} on the left-hand side $P$ behaves like $x^2$ while on the right-hand side, the density term $\varrho^{4/3}$ behaves like $x^{4/3}$. To be consistent, the square bracket in Eq.~\ref{pgpr} must behave like $x^{2/3}$ \citep{Chandra39}, i.e.
 \begin{equation}
  \left[\frac{3}{a} \left(\frac{k}{\mu \, m_{\mathrm{u}}}\right)^4
\, \frac{1-\beta}{\beta^4} \right]^{\frac{1}{3}}  \, \sim \, M^{2/3} \; .
\end{equation}
This means that 
\begin{equation}
\frac{(1-\beta)^{1/2}}{\beta^2} \, \sim  \, \, \mu^2 \, M  \; .
\label{functionBmu2}
\end{equation}
Thus, we have returned to an equation similar to Eq.~\ref{functionBmu},  which applies to any point in the star, including the centre.\\

\noindent
The general dependence of $\beta$ on $\mu^2  M$ can also be verified by expressing the ratio of the radiation to gas pressures
\begin{equation}
 \frac{P_{\mathrm{rad}}}{P_{\mathrm{gas}}} \, = \, \frac{1}{3}\frac{a \, T^3}{\varrho }\,
 \frac{\mu  m_{\mathrm{u}}}{k}  \; .
\end{equation}
 We take  the average density and   average temperature ${T} \,  = \, \frac{\beta}{3} \, \frac{\mu  m_{\mathrm{u}}}{k} \,q \, \frac{G \, M}{R} $, where $q$ is a concentration factor, with $q=3/5$ for an homogeneous density and $q=3/2$ for a polytrope with $n=3$. The above ratio becomes
 \begin{equation}
 \frac{P_{\mathrm{rad}}}{P_{\mathrm{gas}}} \,\approx \,\frac{4\, \pi}{3^5} \, a \, 
 \left(\frac{m_{\mathrm{u}}}{k}\right)^4 \,q^3 \, G^3 \,  
 \beta^3 \, \mu^4  M^2  \; .
 \end{equation}
 Since $P_{\mathrm{rad}}/P_{\mathrm{gas}}=(1-\beta)/{\beta} $, one gets finally \citep{MaederBook09}
 
\begin{equation}
 \frac{\beta^2}{(1-\beta)^{1/2}} \;  \mu^2  M  \, \approx \,  \left(\frac{3^5}{4 \, \pi \, a}\right)^{1/2} \left(\frac{k}
 {m_{\mathrm{u}}} \right) ^2 \, \frac{1}{G^{3/2}} \frac{1}{q^{3/2}} \; .
 \label{pradm}
 \end{equation}
There are only constant terms in the second member. For $q=3/5$ (homogeneous density), the second member is equal to 21.93 $M_{\sun}$  and for $q=3/2$ (polytrope $n=3$), it is equal to  5.55 $M_{\sun}$.Thus, we have verified in two ways that $\beta$ is a function only of $\mu^2  M $. As mentioned above, this necessarily implies that for some range of $\Gamma$--values smaller than 1.0, the term $\mu^2 M$ is bounded by a constant.

We estimate this constant,  for example, for a 120 $M_{\sun}$ star on the ZAMS with $X=.68$, $Z=0.02$, $\mu=0.627$, $\log (L/L_{\sun} )=6.252$, central and average densities of 1.483 and 0.0438 g cm$^{-3}$, and $\beta=0.495$ \citep{MaederBook09}. These parameters give on the left side of Eq.~\ref{mucarre1} a value $\mu^2 M = 47.18$ $M_{\sun}$, while on the right side we have
 \begin{equation}
 \left( \frac{4 \pi \,c  \,G }{ A} \right)^{1/2}\, \left(\frac{\rho_\mathrm{c}}{\rho}\right)^2
 \, \left( \frac{1}{\beta_{\mathrm{c}}} \right)^2 \, = \,  75.98 \;M_{\sun} \, .
 \end{equation}
 \noindent
The equality of the two sides of Eq.~\ref{mucarre1} is realised for $\Gamma=0.386$. This value is  consistent with the $\Gamma$-value of a 120 $M_{\sun}$ in Fig.~\ref{eddzams}. This means that if this $\Gamma$-value were to be an effective limiting value, the limit $\mu^2 M = 47.18$ $M_{\sun}$ would not be overcome. For a pure helium star, the corresponding limit $M_{\Gamma}$, for this particular $\Gamma$, would be 26.55 $M_{\sun}$ instead of 120 $M_{\sun}$ on the ZAMS.

\begin{figure}[t]
\begin{center}
\includegraphics[height=8.5cm,width=9.0cm]{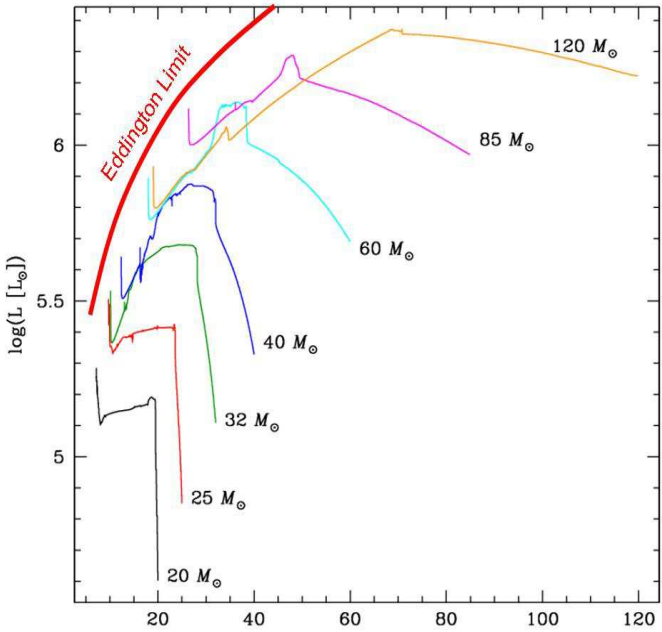}
\caption{Mass-luminosity relations of evolutionary  models with initial $X=0.72$ and $Z=0.014$ for massive stars with an initial rotational velocity equal to 40\% of the critical velocity \citep[from][]{Ekstrom2012a}.}
\label{EDDWR}
\end{center}
\end{figure}

As another example, we see from Figs. \ref{eddzams} and \ref{eddwr} that, for $\Gamma=0.25$ (where the intersection of the $\Gamma$--line with the M--L relation is well-defined), the ratio of the two masses 57 $M_{\sun}$ (with $\mu=0.617$) and 12 $M_{\sun}$ (with $\mu=1.340$ ) corresponds  to the inverse of their $\mu^2$--values. 

A crucial question now is whether during some phase the WR stars evolve keeping the same value of $\Gamma$. Fig. \ref{EDDWR} shows the mass-luminosity relation for rotating stars of different masses evolving away from the ZAMS \citep{Ekstrom2012a}. The evolution is followed until the end of the central C-burning phase. The location of the Eddington limit for stars without hydrogen is indicated. We see that  rather long parts of the evolutionary tracks in the phases of WR stars  without hydrogen (types WNE and WC) are parallel to the Eddington limit at the corresponding masses. These tracks typically have a luminosity that is lower than the Eddington limit by a factor of about two. They correspond to a line $\Gamma \approx 0.5$, (the track for 85 $M_{\sun}$  corresponds to $\Gamma = 0.6$, but is still parallel).

Since these tracks have been calculated with mass-loss rates based on observations, the aforementioned parallelism indicates that the $\Gamma$-factors of WR stars remain constant  during the concerned parts of the evolution. Incidentally, we note that the above value of $\Gamma=0.5$ is an upper bound to the values $\Gamma= 0.20$--0.45 observed in the Arches cluster \citep{graefener11}. This is consistent with the stellar models of Fig.~\ref{EDDWR} seeming to predict too high WR luminosities with respect to the observed ones. This may be an indication that the mass-loss rates applied in the models are too low at some stages \citep{Georgy2012b}.

As a conclusion of the last two sections, we note that, in a sequence of homologous chemically homogeneous WR models, the product $\mu^2 \, M$ is effectively bounded by a constant factor
\begin{equation}
\mu^2  \frac{M}{M_{\sun}} \, \leq   \, C_{\mathrm{EDD}} \; ,
\label{cond}
\end{equation}
where we  call $ C_{\mathrm{EDD}}$ this constant. In the next section, we try to estimate the value of $C_{\mathrm{EDD}}$ from observations of WR masses in the solar neighborhood. 

Here, it is very important to realise that Eq.~\ref{mucarre1} and consequently Eq.~\ref{cond} are just logical consequences of the classical Eddington limit  for electron-scattering opacity. Thus, the above results do not contain subtle consequences of physical effects that are not present in the hypotheses. For example, if we considered such an effect as the metallicity dependence of the WR mass-loss rates and if this effect were to arise from bound-free and line opacities, it would be normal that our results do not contain it. Similarly, if some effects in the mass-loss rates originated from rotation, magnetic field, or binarity, it should also not be expected that the present results reproduce them. To summarise, our results are just equivalent forms of the classical Eddington limit based on electron-scattering opacity for some specific assumptions. In particular, Eqs.~\ref{mucarre1} or \ref{cond} only apply to polytropes of a given n-index and to evolutionary sequences of chemically homogenous stars evolving homologously. For cases that significantly deviate from these hypotheses, our conclusions do not apply.

Therefore, $C_{\mathrm{EDD}}$ is not a "universal constant" for all stars. For example, evolutionary sequences of different metallicity $Z$ are likely to have different values of $C_{\mathrm{EDD}}$ for two reasons. Firstly, the product $\left( {4 \pi \,c  \,G }/{ A} \right)^{1/2}\, \left({\rho_\mathrm{c}}/{\rho}\right)^2 \, \left( {1}/{\beta_{\mathrm{c}}} \right)^2 \,$ may be slightly different. Secondly, the effective $\Gamma$ expressing the ratio of the real observed maximum luminosity to the theoretical Eddington luminosity for electron-scattering opacity may be different as discussed in Sect. 2.3. The same kind of remark may apply to other differences of input parameters. Within homologous sequences of homogeneous models with different input parameters, each sequence may evolve with $\mu^2 \, M= \text{constant}$, but with different values for the constant. However, polytropes with $n=3$ all have the same constant as shown in Sect. 3, thus departures from this polytropic relation is an interesting point to check.
 
Incidentally, we remark in  Fig. \ref{EDDWR} that the strong brightening of the WR stars in the very late stages (from the core contraction that precedes central C burning) brings the star very close to the Eddington limit. This suggests that the WR stars  may experience extremely high mass-loss rates in the relatively fast stages preceding the supernova explosion. When this strong mass loss occurs, just before the SN explosion, it may hide from view the progenitors embedded in a dense envelope. This may be the reason why \citet{Smartt09araa} reports no detection of progenitors for ten Ibc supernovae having pre-explosive images.

\section{A quantitative estimate based on WR stars\label{SecObs}}

For homogeneous stars near the Eddington limit such as WR stars \citep{graefener11}, the condition in Eq.~\ref{cond} implies that the rate of mass loss d$M/$d$t$ is determined by the growth rate of the mean molecular weight $\mu$ resulting from the internal nuclear reactions
\begin{equation}
\frac{\mathrm{d} M}{\mathrm{d} t} \; = \; - 2\, \frac{M}{\mu}  \frac{\mathrm{d} \mu}{\mathrm{d} t} \; .
\label{perteM}
\end{equation}
Interestingly enough, we see that the exact value of $ C_{\mathrm{EDD}}$  does not play a role in the amplitude of the mass-loss rates.

However, the value of $C_{\mathrm{EDD}}$ is still important to determine at which stage of  evolution the condition in Eq.~\ref{cond} and rates Eq.~\ref{perteM} start to apply. For a star on the ZAMS, the value of the product $\mu^2 M$ may initially be below the limit $C_{\mathrm{EDD}}$. However, as a result of nuclear reactions the value of $\mu$ may increase up to a point where the product $\mu^2 M$ reaches the limiting $C_{\mathrm{EDD}}$. From there, in the sequence of homogeneous homologous models, the mass-loss rates will be controlled by Eq.~\ref{perteM}. This does not necessarily mean that, for a given WR mass, the mass-loss rates will always be the same, since depending on  $ C_{\mathrm{EDD}}$ the relation between $M$ and $\mu$ will not be the same and the rate of change  ${\mathrm{d} \mu}/{\mathrm{d} t}$ may also be different.

To estimate the effective constant $C_{\mathrm{EDD}}$,  we now consider various groups of stars at different stages of evolution, studied in most recent literature. We first consider O-type stars on the ZAMS, which evidently are homogeneous with typically a $\mu$--value of about 0.62. On the basis of a large sample  of  90 clusters, \citet{masch08} determined the masses of the most massive stars, to examine whether these are related to the cluster mass. The most massive star they found has a mass of  80 $M_{\sun}$. This is lower than the extreme cases of some LBV and H-rich WN stars, which have a luminosity corresponding to masses of up to 120 $M_{\sun}$ or above \citep{HumphSta05}. On the other hand, the lowest initial stellar  masses leading to WR stars are around 25 $M_{\sun}$ \citep{Maeder81c,Crowther07,sander2011}. Thus, for ZAMS stars we consider that the progenitors of WR stars lie in the range of from 25 $M_{\sun}$ to about 120 $M_{\sun}$, most of them originating from stars with initial masses below 80 $M_{\sun}$.

A most useful constraint comes from the WN stars without evidence of hydrogen at their surface. These stars are essentially pure helium stars with a mean molecular weight of 4/3, although some of them might already have both C and O deep in their interior that has not yet reached the surface. However, at this stage the products of He burning would relatively soon appear in the outer layers. Thus, it is likely that these WN stars have a structure close to chemical homogeneity with  an average $\mu$--value not far from 4/3. A large sample of WN stars without hydrogen has been analysed in the study  by \citet{HamGraf06}. We found that 30 of them have actual masses between 13 $M_{\sun}$ and 35 $M_{\sun}$ with an average of 18.5 $M_{\sun}$. There is only one star with a mass above 25 M$_{\sun}$, thus most are in the range of from 13 $M_{\sun}$ to 25 $M_{\sun}$. Masses of WN stars in binary systems were also determined. The lower bound is a value of 10 $M_{\sun}$, and the upper bound is not meaningful since large WN stars of large masses are generally hydrogen rich \citep{Crowther07}. Thus, WN binaries have about the same lower mass as in the data of \citeauthor{HamGraf06} with a marginal trend to be slightly lower.

\begin{figure}[t] 
\begin{center}
\includegraphics[width=9.2cm]{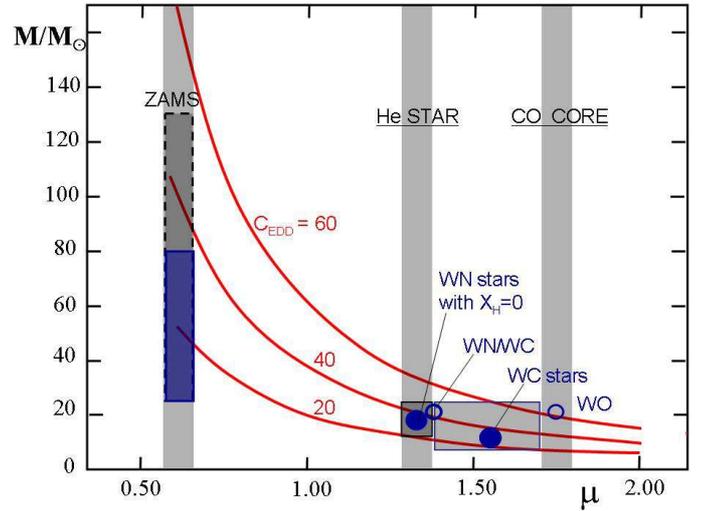}
\caption{The curves $\mu^2 M\,= \,C_{\mathrm{EDD}}$ for 3 values of $C_{\mathrm{EDD}}$. The location of the ZAMS, He-stars and CO-cores are indicated. The average point for 30 WN stars without hydrogen studied by \citet{HamGraf06} is shown, with indications of the limits within they are observed. From the study by \citet{sander2011}, we show the averages for 4 transition WN/WC stars, for 39 WC stars (with indications of the limits) and for 2 WO stars.}
\label{cedd}
\end{center}
\end{figure}
Another interesting constraint is provided by WC stars. These are essentially stars in the process of helium burning, that is to say they are helium stars with already a certain amount of C and a smaller amount of O. In view of the lifetimes of the nuclear burning phases, only  about 1\% of WC stars may already be in the stage of central C burning and develop strong chemical  inhomogeneity. Thus, most of the WC stars are likely to have a $\mu$--value between that of ionised helium, i.e. 1.33 and that of ionised carbon, i.e. 1.71 (for ionised oxygen $\mu= 1.78$). There may be some departures from a constant internal mean molecular weight in the WC phase. However, these are probably not large in view of the heavy mass loss and the relatively limited range of $\mu$--values present. The stellar parameters of 39 WC stars were determined by \citet{sander2011}. The  average actual mass, as determined from the luminosities, is 12.9 $M_{\sun}$. This is quite consistent with the values of the masses of WC stars determined from binary WR stars. These masses range from 9 to 16 $M_{\sun}$ \citep{Crowther07}. There are also four transition WN/WC stars with an average mass of 21.5 $M_{\sun}$ and two WO stars with an average mass of 22 $M_{\sun}$. As pointed out by \citet{sander2011},  the WO stars may stem from the most massive stars. Their location in Fig.~\ref{cedd} would agree with this conclusion.

These various data are represented in Fig.~\ref{cedd}, which shows masses versus mean molecular weights. We indicate the curves corresponding to $C_{\mathrm{EDD}}$= 20, 40, and 60. We see that the data based on significantly large samples are consistent with a general average value of $C_{\mathrm{EDD}}$ between 20 and 40, if we account for the weight of the  samples of WN and WC stars. It is indeed uncertain whether the evolutionary sequence from O-stars to WN and WC stars forms a unique homologous sequence. This path may be "tangential" to one or two different homologous sequences with slightly different values of $C_{\mathrm{EDD}}$. The more the homogeneity is satisfied, the more of course the homology property will be verified. This means that the observationally estimated value of $C_{\mathrm{EDD}}$ between 20 to 40 is an average value covering the evolution from the MS of massive stars to WR stars and that in the future it might perhaps be more clearly specified for some parts of the evolution.

Figure \ref{cedd} also shows that the above observations are consistent with a curve of the type given by Eq.~$\ref{cond}$, although the WR stars originate from a rather broad interval of initial masses. These results are in agreement with the mass-loss rates of WR stars being governed by the proximity to the Eddington limit, as stated  by \citet{graefener11}.

\section{Conclusions\label{SecConclu}}

We have shown that, in a  sequence of homologous, chemically homogeneous stars, a relation of the form of Eq.~\ref{cond} is followed, if the star keeps a constant $\Gamma$-factor. Different homologous sequences characterised by different input parameters may have different values of the constant in Eq.~\ref{cond}. Recent determinations of WR masses of various subtypes tend to support an evolution according to the condition given in Eq.~\ref{cond} and allow us to estimate that the average value of the effective numerical constant in Eq.~\ref{cond} is between 20 and 40.

The validity of our aforementioned hypotheses (homogeneity, homology, constant $\Gamma$) must of course be verified in the domain where the above relation would be  applied in evolutionary calculations. 

\begin{acknowledgements}
The authors thank an anonymous referee for constructive remarks.
\end{acknowledgements}

\bibliographystyle{aa}
\bibliography{BibTexRefs}

\end{document}